\documentclass[aps, pra, twocolumn, amsmath, amssymb, superscriptaddress, nofootinbib]{revtex4-1}

\usepackage[utf8]{inputenc}
\usepackage{graphicx}
\usepackage{units}
\usepackage{color}
\usepackage[pdftex, colorlinks=true, linkcolor=myblue, citecolor=myblue,
urlcolor=myblue]{hyperref}
\usepackage{times}
\usepackage{comment}
\usepackage{graphicx}
\usepackage{amsmath, calc}
\usepackage{mathrsfs}
\usepackage{amsfonts}
\usepackage{amssymb}
\usepackage{bm}
\usepackage{bbm}
\usepackage{color}
\usepackage{array}
\usepackage{units}

\definecolor{myblue}{rgb}{0,0,1}
\definecolor{myred}{rgb}{1,0,0}

\begin{document}
%\narrowtext
\title{Topological phases of polaritons in a cavity waveguide}

\author{C.~A.~Downing}
\affiliation{Departamento de F\'{i}sica de la Materia Condensada, CSIC-Universidad de Zaragoza, Zaragoza E-50009, Spain}

\author{T.~J.~Sturges}
\affiliation{Institute of Theoretical Physics, Faculty of Physics, University of Warsaw, ul.~Pasteura 5, 02-093, Warsaw, Poland}

\author{G. Weick}
\affiliation{Universit\'{e} de Strasbourg, CNRS, Institut de Physique et Chimie des Mat\'{e}riaux de Strasbourg, UMR 7504, F-67000 Strasbourg, France}

\author{M.~Stobi\'{n}ska} 
\affiliation{Institute of Theoretical Physics, Faculty of Physics, University of Warsaw, ul.~Pasteura 5, 02-093, Warsaw, Poland}

\author{L.~Mart\'{i}n-Moreno}
\email{lmm@unizar.es} 
\affiliation{Departamento de F\'{i}sica de la Materia Condensada, CSIC-Universidad de Zaragoza, Zaragoza E-50009, Spain}

%===========================================================================
%===========================================================================
%===========================================================================
%===========================================================================
%===========================================================================
%===========================================================================
%===========================================================================
%===========================================================================

\begin{abstract}

We study the unconventional topological phases of polaritons inside a cavity waveguide, demonstrating how strong light-matter coupling leads to a breakdown of the bulk-edge correspondence. We observe an ostensibly topologically nontrivial phase, which unexpectedly does not exhibit edge states. Our findings are in direct contrast to topological tight-binding models with electrons, such as the celebrated Su-Schrieffer-Heeger (SSH) model. We present a theory of collective polaritonic excitations in a dimerized chain of oscillating dipoles embedded inside a photonic cavity. The added degree of freedom from the cavity photons upgrades the system from a typical SSH $\mathrm{SU}(2)$ model into a largely unexplored $\mathrm{SU}(3)$ model. Tuning the light-matter coupling strength by changing the cavity size reveals three critical points in parameter space: when the polariton band gap closes, when the Zak phase changes from being trivial to nontrivial, and when the edge state is lost. These three critical points do not coincide, and thus the Zak phase is no longer an indicator of the presence of edge states. Our discoveries demonstrate some remarkable properties of topological matter when strongly coupled to light, and could be important for the growing field of topological nanophotonics.
\end{abstract}

%===========================================================================
%===========================================================================
%===========================================================================
%===========================================================================
%===========================================================================
%===========================================================================
%===========================================================================
%===========================================================================

\maketitle

%\pacs{73.22.Pr, 73.21.La, 03.65.Ge, 03.65.Pm}

%===========================================================================
%===========================================================================
%===========================================================================
%===========================================================================
%===========================================================================
%===========================================================================
%===========================================================================
%===========================================================================

\textit{Introduction}.--The utility of topology in quantum materials has been proven by the revolutions in the understanding of a range of phenomena, including the quantum Hall effect and the spin-liquid states of antiferromagnets \cite{Thouless1998, Kosterlitz2017, Haldane2017}. Recently, a rapidly growing research field known as topological nanophotonics has emerged \cite{Ozawa2019, Rider2019}, which seeks to harness the power of topology in a new generation of controllable light-based structures \cite{Lu2014, Khanikaev2017, Sun2017}. It is envisaged that this topic will lead to a deeper understanding of light-matter interactions at a fundamental level \cite{Martinez2018, Ghatak2019}, as well as novel applications including chiral lasers \cite{Jean2017}, integrated quantum optical circuits \cite{Barik2018}, and nonlinear light generators \cite{Kruk2019}.

The prototypical one-dimensional (1-D) model exhibiting nontrivial topology is the Su-Schrieffer-Heeger (SSH) chain \cite{Su1979, Su1980}. Fundamentally, this model is a tight-binding Hamiltonian with nearest neighbor hopping in a dimerized 1-D lattice \cite{Bernevig2013, Asboth2016, Cooper2019}. In its topologically nontrivial phase, which is associated with a Zak phase of $\pi$ \cite{Berry1984, Zak1989}, it hosts edge states at the ends of the chain. This accordance between the Zak phase (a band dependent quantity) and the presence of edge states is known as the bulk-edge correspondence \cite{Asboth2016, Cooper2019}. The SSH model has recently been exploited in a series of works, including in photonics \cite{Schomerus2013, Slobo2015}, plasmonics \cite{Ling2015, Downing2017, Pocock2018, Downing2018}, polaritonics \cite{Solny2016, Solny2016b}, superconducting circuits \cite{Engelhardt2017}, and waveguide QED \cite{Bello2019}. Cutting edge experiments have begun to report some remarkable topological properties of extended SSH models in these research fields \cite{Sinev2015, Xiao2017, Parto2018, Whittaker2019, Han2019}.

We are concerned with an extension of the SSH model, staged by polariton excitations \cite{Hopfield1958, Torma2015} inside a cavity. We are interested in how the intrinsic topology of the SSH model, whose role is played by collective excitations coupled throughout a dimerized chain, is changed in novel directions due to strong light-matter coupling \cite{Ameling2013, Ginzburg2016}. Modulating the cavity dimensions allows one to tune the light-matter coupling strength and the detuning between the dipolar excitations and cavity photons, allowing us to provide insight into the breakdown of the bulk-edge correspondence. Technically, one can adjust the system from a standard SSH $\mathrm{SU}(2)$ model (resembling the Peierls chain \cite{Peierls1991}) into a reconstructed $\mathrm{SU}(3)$ system, composed of polaritons formed from the coupling of the two SSH bands and a cavity photonic band. This has profound consequences for the topological phases of our system of topological polaritons, which are noticeably very different from the ``topolaritons'' generated via a winding exciton-photon coupling \cite{Karzig2015, Bardyn2015, Yi2016, Gutierrez2018}. Usually, there is one critical point in parameter space for SSH models: where the Zak phase changes from nontrivial to trivial, the edge states are lost, and the band gap closes \cite{Bernevig2013, Cooper2019, Delplace2011}. We find three different points in parameter space at which these transitions happen, in a manifestation of a breakdown of the bulk-edge correspondence. There is an apparently topologically nontrivial phase, as codified by the Zak phase, formally without any edge states. We observe the remnants of edge states in the continuous part of the spectrum, rather than in the band gap, which we term ``mixed states''. We note that our system is Hermitian, and as such is different from the breakdowns discussed in non-Hermitian models \cite{Kunst2018, Yao2018, Pocock2019, Borgnia2019}. 

Our proof-of-principle model of the bulk-edge correspondence breakdown disregards losses, since we are interested in constructing a minimal Hermitian system which can capture this novel physics. As such, realizing our model of a dipolar chain via the Mie resonances of dielectric nanoparticles is experimentally appealing \cite{Slobo2015, Du2011, Bakker2017}. Metallic nanoparticles hosting plasmonic excitations are an alternative platform which correspond to our theory, and which benefits from extensive study \cite{Brongersma2000, Maier2003, Park2004, Weber2004, Citrin2004, Simovski2005, Koenderink2006, Markel2007, Compaijen2013, Petrov2015, Brandstetter2016, Downing2018b, Compaijen2018}. Our model is in fact relevant for any system where dipolar interactions are the dominant mechanism, for example microwave helical resonators \cite{Mann2018}, magnonic  microspheres \cite{Pirm2018}, and Rydberg \cite{Browaeys2016, Leseleuc2019} and cold atoms \cite{Perczel2017}. 

%===========================================================================
%===========================================================================
%===========================================================================
%===========================================================================
%===========================================================================
%===========================================================================
%===========================================================================
%===========================================================================

\begin{figure}[tb]
 \includegraphics[width=1.0\linewidth]{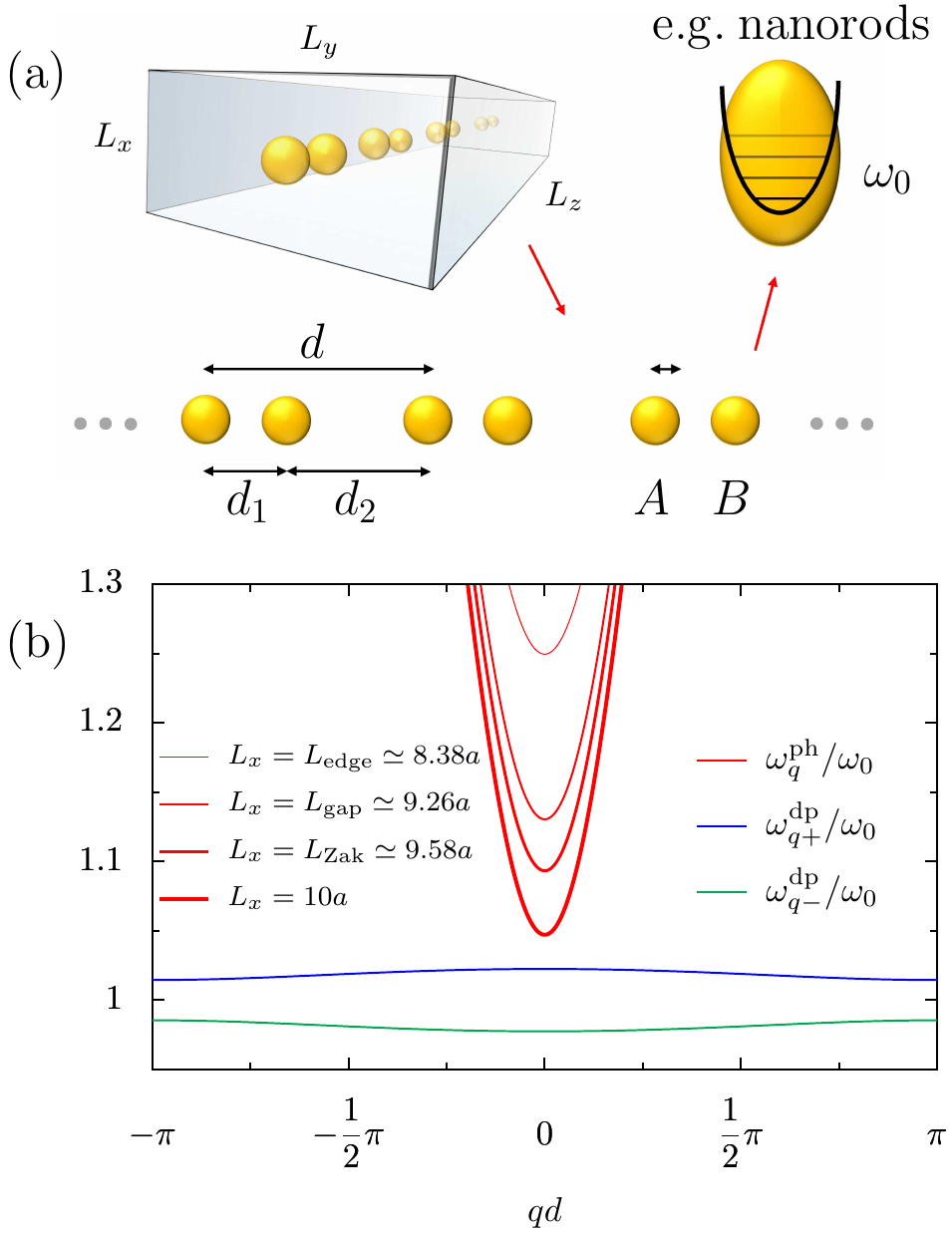}
 \caption{(a) Schematic of our system: a cuboid cavity encasing a dimerized chain (lattice constant $d$) of dipolar excitations (resonance frequency $\omega_0$ and length scale $a$), which can be realized by, e.g., nanorods. Light-matter interactions induce nontrivial topological phases of the resultant polariton excitations. (b) Cavity photon dispersion $\omega_q^{\mathrm{ph}}$ for several values of cavity height $L_x$, with the aspect ratio $L_y = 3 L_x$. As the cross-sectional area increases (thicker lines) the photon band approaches the SSH bands $\omega_{q \pm}^{\mathrm{dp}}$. In panel (b), $d = 8a$, $\varepsilon = 0.25$, and $\omega_0 a/c = 0.1$.}
 \label{sketch}
\end{figure}

\textit{Dipolar chain}.--We consider a dimerized chain of oscillating dipoles in the form of an SSH lattice, as sketched in Fig.~\ref{sketch}~(a). The unit cell of the chain is comprised of a dimer of dipoles (labeled $A$ and $B$), which introduces two length scales: the alternating center-to-center distances $d_1$ and $d_2$. The chain is characterized by the lattice constant $d = d_1 + d_2$, and the dimerization parameter $\varepsilon = (d_1-d_2)/d$. Each dipole corresponds to a bosonic excitation with a resonance frequency $\omega_0$ \cite{SupplementalMaterial}. Dipolar interactions among the localized excitations causes the formation of collective modes throughout the chain. In this setup, and restricting the coupling to nearest-neighbors \cite{SupplementalMaterial}, the collective dipolar ($\mathrm{dp}$) excitations mimic the structure of the SSH Hamiltonian \cite{Ling2015, Downing2017, Pocock2018, Downing2018}
\begin{equation}
\label{eq:Ham_two_band}
 \mathcal{H}_{\mathrm{dp}} = \begin{pmatrix}
\omega_{0}   && \Omega g \\ 
\Omega g^{\ast}  && \omega_{0}
 \end{pmatrix}, \quad g = (d/d_1)^3 + (d/d_2)^3 \mathrm{e}^{-\mathrm{i} qd}.
 \end{equation}
The coupling constant $\Omega = \omega_0 (a/d)^3 /2$, where $a$ is a length scale characterizing the strength of the dipolar excitations. The resulting collective dipolar excitation bandstructure $\omega_{q \pm}^{\mathrm{dp}} = \omega_0 \pm \Omega |g|$ is the SSH dispersion, and is shown in Fig.~\ref{sketch}~(b) as a function of the wavevector $q$. Here the upper $+$ (lower $-$) band, as drawn by the solid blue (green) line, is bright (dark). This is because the dipole moments in a given dimer are in-phase $\uparrow \uparrow$ (out-of phase $\uparrow \downarrow$), such that the coupling to light is significant (negligible) \cite{Downing2017, Downing2018}. When $\varepsilon > 0~(\varepsilon < 0)$ the Zak phase is $\pi~(0)$, corresponding to the presence (absence) of edge states in a finite chain \cite{Bernevig2013, Asboth2016, Cooper2019}.

\textit{Cavity photons}.--The dipolar chain is embedded in a cuboid cavity of volume $L_x \times L_y \times L_z$ [see Fig.~\ref{sketch}~(a)], whose structure quantizes the electromagnetic field into photonic bands \cite{Milonni1994, Kakazu1994}. For a long cavity $L_z \gg L_x, L_y$, only a single photonic band $\omega_q^{\mathrm{ph}} = c \sqrt{q^2 + (\pi/L_y)^2}$, where $c$ is the speed of light in vacuum, is relevant due to its proximity to the resonance frequency $\omega_0$ \cite{SupplementalMaterial}. We fix the aspect ratio $L_y = 3 L_x$ throughout, and in Fig.~\ref{sketch}~(b) we show how the photonic band may be lowered towards the SSH bands by increasing the cavity cross-section (as shown by the increasingly thick red solid lines). We show three named length scales $\{ L_{\mathrm{edge}}$, $L_{\mathrm{gap}}, L_{\mathrm{Zak}} \}$, whose importance for the topological phases will be explained later. In Fig.~\ref{sketch}~(b), we have displayed the uncoupled SSH and photonic bands. However, light-matter interactions will necessarily see these bands hybridize into polaritons \cite{Weick2015, Lamowski2018, Mann2018}, changing the resulting bandstructure.

%===========================================================================
%===========================================================================
%===========================================================================
%===========================================================================
%===========================================================================
%===========================================================================
%===========================================================================
%===========================================================================

\begin{figure*}[tb]
 \includegraphics[width=1.0\linewidth]{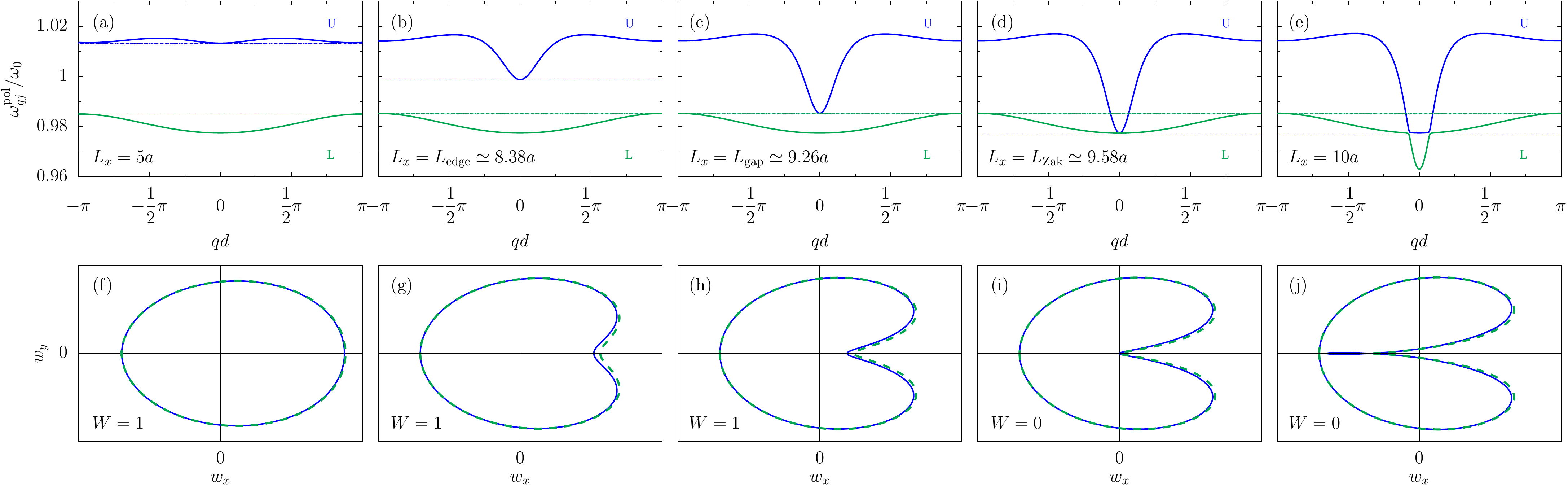}
 \caption{ Upper panels: polariton bandstructure $\omega_{q j}^{\mathrm{pol}}$, with increasing cavity size from (a) to (e) (cavity height $L_x$ increases and we set $L_y = 3 L_x$). The thin dashed lines show the minimum (maximum) of the $j= \mathrm{U} (\mathrm{L})$ band. The (mostly photonic) $\mathrm{P}$ band is not visible on this scale (see Sec.~II of Ref.~\cite{SupplementalMaterial} for more details). Lower panels: winding vector ($w_x, w_y$) of the effective $\mathrm{SU}(2)$ Hamiltonian $\boldsymbol{\sigma} \cdot \boldsymbol{w}$, with (f-j) corresponding to the panels (a-e) directly above. Results for the band $j= \mathrm{U} (\mathrm{L})$ are denoted by blue (green) lines. Same parameters as in Fig.~\ref{sketch}. }
 \label{continuum}
\end{figure*}

\textit{Polariton model}.--The light-matter coupling modifies the Hamiltonian of Eq.~\eqref{eq:Ham_two_band} from a $\mathrm{SU}(2)$ SSH model into an $\mathrm{SU}(3)$ polaritonic model, which for a given $q$ reads~\cite{SupplementalMaterial}
\begin{equation}
\label{eq:Ham_three_band}
    \mathcal{H}_{\mathrm{pol}} = \left(\begin{array}{cc|c}
        &         &     \mathrm{i} \xi \mathrm{e}^{-\mathrm{i} \chi}  \\
      \multicolumn{2}{c|}{\smash{\raisebox{.5\normalbaselineskip}{$\mathcal{H}_\mathrm{dp}$}}} & \mathrm{i} \xi \mathrm{e}^{\mathrm{i} \chi} \\
      \hline \\[-\normalbaselineskip]
      -\mathrm{i} \xi \mathrm{e}^{\mathrm{i} \chi}  & -\mathrm{i} \xi \mathrm{e}^{-\mathrm{i} \chi} & \omega_{q}^{\mathrm{ph}}
    \end{array}\right).
\end{equation}
$\mathcal{H}_{\mathrm{dp}}$ is defined in Eq.~\eqref{eq:Ham_two_band}, $\xi = (2 \pi a^3 \omega_0^3/ d L_x L_y \omega_q^{\mathrm{ph}} )^{1/2}$ is the light-matter coupling constant, and $\chi = q d_1/2$ arises because the light-matter coupling is evaluated at the two inequivalent lattice sites $A$ and $B$. In the upper panels of Fig.~\ref{continuum}, we plot the resulting polariton bandstructure $\omega_{q j}^{\mathrm{pol}}$, where the band index $j = \{ \mathrm{P}, \mathrm{U}, \mathrm{L} \}$ labels (for a given wavevector $q$) the eigenfrequencies in descending order. The mostly photonic $\mathrm{P}$ band is not visible on this scale as it is higher in frequency, leaving one to view the two lowest bands (denoted by $\mathrm{U}$ and $\mathrm{L}$, solid blue and green lines, respectively), since we are principally interested in the polariton bands $\omega_{q \mathrm{U} / \mathrm{L}}^{\mathrm{pol}}$ which are those smoothly deformed from the dipolar bands $\omega_{q \pm}^{\mathrm{dp}}$. In going from panel (a) to (e), we increase the cavity cross-sectional area, which reduces the light-matter detuning and thus causes significant reconstructions of the standard SSH dispersion [cf. Fig.~\ref{sketch}~(b)]. The thin dashed lines show the minimum of the $\mathrm{U}$ (blue) and $\mathrm{L}$ (green) bands, in order to highlight the behavior of the polaritonic band gap, which we define as $\mathrm{min} \{ \omega_{q \mathrm{U}}^{\mathrm{pol}} \} - \mathrm{max} \{ \omega_{q \mathrm{L}}^{\mathrm{pol}} \}$. The latter quantity is reduced in going from panel (a) to (b), until it is finally zero in panel (c) at the critical length $L_x = L_{\mathrm{gap}}$ ($L_y = 3 L_x$). Further increase of the cavity area leads to a negative band gap and panel (d), which is associated with the critical length $L_x = L_{\mathrm{Zak}}$. This corresponds to the critical point at which the Zak phase changes from being nontrivial to trivial (in the SSH model this would correspond to the loss of edge states) \cite{Asboth2016, Delplace2011}. Thus the changing of the Zak phase and closing of the global band gap no longer coincide. In going from panel (d) to (e), one notices that for even greater cavity sizes the system displays the band anti-crossing phenomenon, such that the negative band gap is essentially constant.

%===========================================================================
%===========================================================================
%===========================================================================
%===========================================================================
%===========================================================================
%===========================================================================
%===========================================================================
%===========================================================================

\begin{figure}[tb]
 \includegraphics[width=1.0\linewidth]{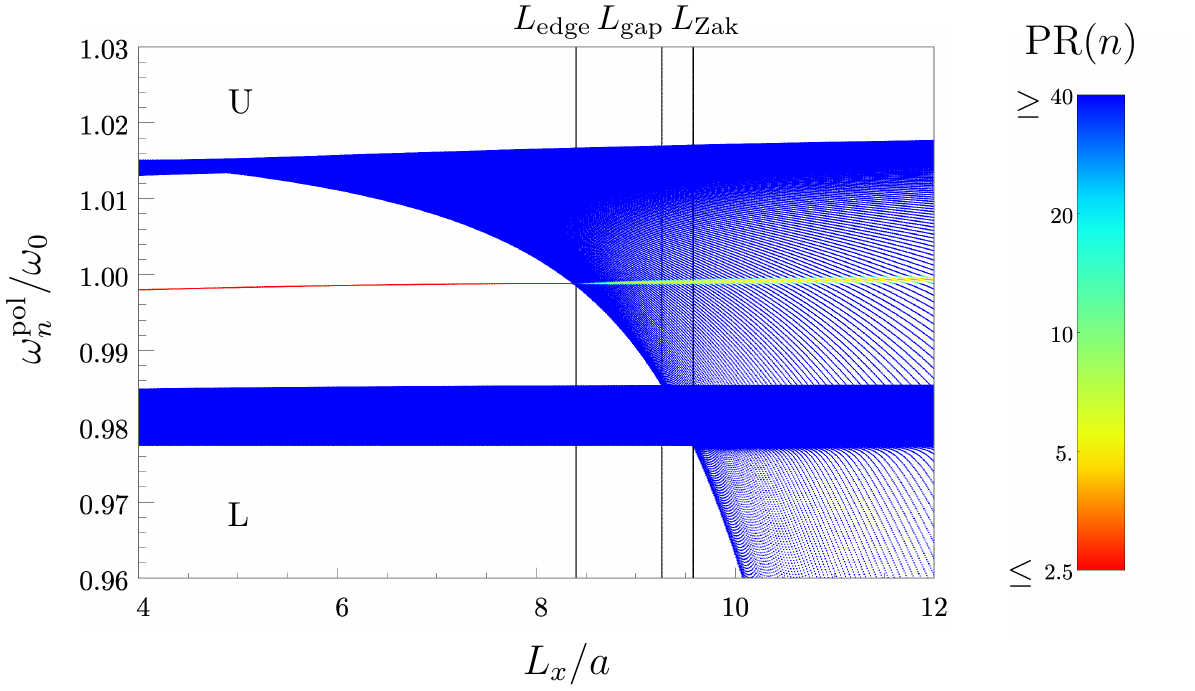}
 \caption{ Polariton eigenfrequencies $\omega_{n}^{\mathrm{pol}}$ in real space as a function of the cavity size (we fix $L_y = 3 L_x$). The color bar measures the participation ratio $\mathrm{PR}(n)$ of each eigenstate, which are indexed by $n$. The number of dimers $\mathcal{N} = 500$. We use the same parameters as in Fig~\ref{sketch}. }
 \label{realspace}
\end{figure}

\begin{figure*}[tb]
 \includegraphics[width=1.0\linewidth]{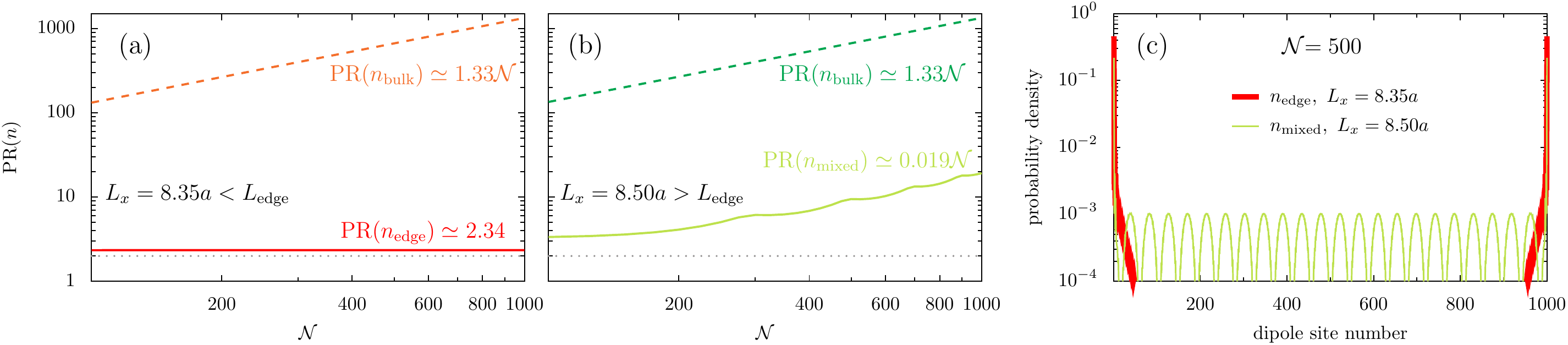}
 \caption{ Panels (a) and (b): participation ratio $\mathrm{PR}(n)$ of a polariton eigenstate, as a function of $\mathcal{N}$. Panel (a): Cavity height $L_x = 8.35 a < L_{\mathrm{edge}}$, and the dashed orange (solid red) line denotes a bulk state (an edge state) associated with the index $n_{\mathrm{bulk}}$ ($n_{\mathrm{edge}}$). Panel (b): Cavity height $L_x = 8.50 a > L_{\mathrm{edge}}$, and the dashed green (solid light green) line denotes a bulk state (a mixed state) associated with the index $n_{\mathrm{bulk}}$ ($n_{\mathrm{mixed}}$). Panel (c): probability density across the dipole sites for a chain of $\mathcal{N} = 500$ dimers, shown for an edge (thick red line) and mixed (thin light green line) eigenstate. Same parameters as in Fig~\ref{sketch}. }
 \label{chainlength}
\end{figure*}

\textit{Topological phases}.--In order to analyze the topological phases and the bulk-edge correspondence, we calculate the polariton eigenfrequencies $\omega_{n}^{\mathrm{pol}}$ in real space (the index $n$ labels each eigenvalue) as a function of the cavity cross-sectional area (see Fig.~\ref{realspace}). Each eigenstate $\psi(n) = (\psi_1, \cdots, \psi_{2 \mathcal{N}} )$ spans each dipole site in the chain of $\mathcal{N}$ dimers, and the color bar measures its participation ratio $\mathrm{PR}(n)$ \cite{Bell1970, Thouless1974}. Defined by $\mathrm{PR}(n) = (\sum_i | \psi_i(n) |^2)^2 / \sum_i | \psi_i(n) |^4$, this quantity gives information about the localization of each state over all of the dipole sites $i$. Edge states have $\mathrm{PR}(n) \simeq 2$ due to being supported only at the two end sites of the chain. For lower cavity heights up to $L_x \simeq 5 a$ in Fig.~\ref{realspace}, we have an effective SSH model with clearly distinct $\mathrm{U}$ and $\mathrm{L}$ bands, corresponding to panel (a) in Fig.~\ref{continuum}. There is an edge state (red line) in the band gap, since the dimerization $\varepsilon > 0$ and there is a Zak phase of $\pi$. The increasing deformation of the polaritonic bands with increasing cavity area sees the edge states dive into the $\mathrm{U}$ polariton band at $L_x = L_{\mathrm{edge}}$. This causes a jump in participation ratio and the formal loss of the edge state (we will see in Fig.~\ref{chainlength} that with even longer chains this jump is even more pronounced). This happens before both the closing of the band gap and the changing of the Zak phase ($L_{\mathrm{edge}} < L_{\mathrm{gap}} < L_{\mathrm{Zak}}$). This is in stark contrast to usual SSH systems and even extended SSH models, for example those with beyond nearest neighbor hopping \cite{Perez2019}. Hence we have observed a complete breakdown of the bulk-edge correspondence, induced by strong light-matter interactions.

It is clear that the second critical point, $L_{\mathrm{gap}}$ in Fig.~\ref{realspace}, shows agreement between continuum and real space diagonalizations, since the $\mathrm{U}$ band is in contact with the $\mathrm{L}$ band at the same point in parameter space [cf. panel (c) in Fig.~\ref{continuum} and the closing of the global band gap]. The real space diagonalization of Fig.~\ref{realspace} also gives a meaning to the Zak phase: while it no longer signifies a loss of edge state, the third critical point $L_{\mathrm{Zak}}$ can be thought of as marking when the $\mathrm{U}$ band first protrudes below the bottom of the $\mathrm{L}$ band, as hinted at by panel (d) in Fig.~\ref{continuum} where the local gap is closed.

%===========================================================================
%===========================================================================
%===========================================================================
%===========================================================================
%===========================================================================
%===========================================================================
%===========================================================================
%===========================================================================

\textit{Winding number}.--An understanding of the bulk-edge correspondence breakdown can be achieved by considering the winding number \cite{Bernevig2013, Asboth2016, Cooper2019}. In the SSH subspace, the two-band system of Eq.~\eqref{eq:Ham_two_band} can be decomposed into $\mathcal{H}_{\mathrm{dp}} = \boldsymbol{\sigma} \cdot \boldsymbol{d}$, where $\boldsymbol{\sigma} = (\sigma_x, \sigma_y)$ is the Pauli vector and $\boldsymbol{d} = (d_x, d_y) = \Omega |g| ( \cos \phi, -\sin \phi)$ is the winding vector, with $\mathrm{e}^{\mathrm{i} \phi } = g/|g|$ \cite{note1}. A parametric plot of the two components of the winding vector in the first Brillouin zone leads to a loop: if this loop encloses the origin the winding number $W = 1$ (corresponding to a Zak phase of $\pi$ and the presence of edge states), and if the loop does not contain the origin the winding number $W = 0$ (a topologically trivial phase with a Zak phase of $0$ and without any edge states) \cite{Asboth2016, Delplace2011}. Our three-band polariton model $\mathcal{H}_{\mathrm{pol}}$ of Eq.~\eqref{eq:Ham_three_band} is associated with $\mathrm{SU}(3)$, frustrating an immediate connection to the graphical representation of topological phases via the winding number $W$. However, one may formulate an exact $\mathrm{SU}(2)$ 3-band polariton Hamiltonian $\mathcal{H}_{\mathrm{pol}} = \boldsymbol{\sigma} \cdot \boldsymbol{w}$, where $\boldsymbol{w} = \Omega |h| ( \cos \varphi, -\sin \varphi)$ and $\mathrm{e}^{\mathrm{i} \varphi } = h/|h|$.  The cost is that the model parameters become polariton frequency dependent, with the role of $g$ now played by the quantity \cite{SupplementalMaterial}
\begin{equation}
\label{eq:g_renorm}
 h = g +  \frac{\mathrm{e}^{- 2 \mathrm{i} \chi}}{\Omega} \frac{\xi^2}{\omega_{q j}^{\mathrm{pol}} - \omega_{q}^{\mathrm{ph}}},
 \end{equation}
which highlights the appearance of self-energy-like terms due to the light-matter interaction. In the SSH limit of our model [cf. panel (a) in Fig.~\ref{continuum} and small $L_x$ in Fig.~\ref{realspace}], the bulk edge correspondence states that the loop spanned by $\boldsymbol{w}$ will enclose the origin ($W=1$) to account for the observed edge state. This is what we obtain, as plotted in panel (f) of Fig.~\ref{continuum}, where the blue and green lines corresponds to the $\mathrm{U}$ and $\mathrm{L}$ bands. As one increases the cavity area in panels (f) to (g) to (h) in Fig.~\ref{continuum}, one sees the novelty of our topological polaritons: a warping of the loop into a structure resembling the lima\c{c}on of Pascal. In the SSH model the structure is an ellipse \cite{Asboth2016, Delplace2011}. In panel (i), when $L_x = L_{\mathrm{Zak}}$, the light-matter interactions have warped the winding vector loop to the extent that it no longer encloses the origin, in agreement with the change in Zak phase, and so $W=0$. However, prior to that, one notices in panels (g) and (h) that neither the loss of edge state nor the closing of the global band gap are marked by any significant feature in the winding vector loop. This analysis highlights that the breakdown of the bulk-edge correspondence is due to the strong light-matter interaction as encapsulated by Eq.~\eqref{eq:g_renorm}, which leaves an $\mathrm{SU}(3)$ fingerprint on the winding vector even when represented in an $\mathrm{SU}(2)$ form.

%===========================================================================
%===========================================================================
%===========================================================================
%===========================================================================
%===========================================================================
%===========================================================================
%===========================================================================
%===========================================================================

\textit{Mixed states}.--We showed in Fig.~\ref{realspace} how the edge states in the system are lost at the critical cavity height $L_{\mathrm{edge}}$, as adjudicated by the jump in participation ratio $\mathrm{PR}(n)$. For smaller heights, edge states are characterized by $\mathrm{PR}(n_{\mathrm{edge}}) \simeq 2$ (dotted gray lines), while bulk eigenstates are described by $\mathrm{PR}(n_{\mathrm{bulk}}) \simeq (4/3) \mathcal{N}$, where $\mathcal{N}$ is the number of dimers in the chain. We show in Fig.~\ref{chainlength}~(a), for a typical cavity size in the sub-critical regime $L_x < L_{\mathrm{edge}}$, how this behavior is robust to increasing the length of the chain. In Fig.~\ref{chainlength}~(b), we present the same information but now for a supercritical cavity height $L_x > L_{\mathrm{edge}}$, illustrating that while the bulk eigenstates continue to be described by $\mathrm{PR}(n_{\mathrm{bulk}}) \simeq (4/3) \mathcal{N}$ (dashed line) the eigenstate with the smallest participation ratio, $\mathrm{PR}(n_{\mathrm{mixed}})$, is no longer an edge state (solid line). Instead, we observe a third type of state -- which we call a mixed state -- since it resides in the continuous part of the spectrum and yet has strong localization (a participation ratio above that of a conventional edge state but below that of the bulk states). In Fig.~\ref{chainlength}~(c), we display the probability density across the dipole sites for the eigenstate associated with the edge state (mixed state) as the thick red (thin green) line [corresponding to $\mathcal{N} = 500$ in panels (a) and (b)]. Contrary to the edge state, the mixed state has a non-negligible contribution in the bulk.

%===========================================================================
%===========================================================================
%===========================================================================
%===========================================================================
%===========================================================================
%===========================================================================
%===========================================================================
%===========================================================================

\textit{Conclusions}.--We have studied the topological phases of polaritons in a 1-D cavity waveguide, showing how strong light-matter coupling leads to a formal breakdown of the bulk-edge correspondence. We have introduced a Hermitian system where, despite being characterized by a nontrivial Zak phase, there are no topological edge states. Our theoretical model can be tuned from an $\mathrm{SU}(2)$ SSH system into an $\mathrm{SU}(3)$ polariton system by changing the cavity size, leading to a number of novelties. The strong coupling regime causes a warping of the winding vector, resulting in a change in the polariton winding number which is out-of-sync with both the loss of edge states and closing of the global band gap. We observe mixed states in the continuum, which are a remnant of the edge states lost with increasing cavity size. This surprising behavior, beyond the SSH model familiar from tight-binding models, shows how strong light-matter coupling necessitates a new theoretical framework in the field of topological nanophotonics.

%===========================================================================
%===========================================================================
%===========================================================================
%===========================================================================
%===========================================================================
%===========================================================================
%===========================================================================
%===========================================================================

We are grateful to \'{A}.\,G\'{o}mez-Le\'{o}n and G.\,Platero for fruitful discussions. CAD acknowledges support from the Juan de la Cierva program (MINECO). TJS and MS were supported by the Foundation for Polish Science ``First Team'' project (No.\,POIR.04.04.00-00-220E/16-00, originally FIRST TEAM/2016-2/17). GW acknowledges financial support from Agence Nationale de la Recherche (Grant No.\,ANR-14-CE26-0005 Q-MetaMat). LMM was supported by the Spanish MINECO (Contract No.\,MAT2017-88358-C3-I-R).

%===========================================================================
%===========================================================================
%===========================================================================
%===========================================================================
%===========================================================================
%===========================================================================
%===========================================================================
%===========================================================================


\begin{thebibliography}{100}
\bibitem{note1}
Here we neglect terms proportional to the identity matrix since they do not affect the form of the eigenstates.
\end{thebibliography}
\end{document}